# Decay Rates of Various Bottomonium systems[*]

Seyong Kim[†]

High Energy Physics Division, Argonne National Laboratory, 9700 South Cass Avenue, Argonne, IL 60439, USA

Using the Bodwin–Braaten–Lepage factorization theorem in heavy quarkonium decay and production processes, we calculated matrix elements associated with S- and P-wave bottomonium decays via lattice QCD simulation methods. In this work, we report preliminary results on the operator matching between the lattice expression and the continuum expression at one loop level. Phenomenological implications are discussed using these preliminary $\overline{MS}$ matrix elements.

Heavy quarkonium decay rates can be written as a sum of factorized products of perturbative coefficients and non-perturbative matrix elements [1],

$$\Gamma(H \to \text{LH}) = \sum_n \frac{2 \text{Im} f_n(\Lambda)}{M^{d_n-4}} \langle H | \mathcal{O}_n(\Lambda) | H \rangle, \quad (1)$$

for the decays into light hadrons and

$$\Gamma(H \to \text{EM}) = \sum_n \frac{2 \text{Im} f_n^{EM}(\Lambda)}{M^{d_n-4}} \langle H | \mathcal{O}_n^{EM}(\Lambda) | 0 \rangle \langle 0 | \mathcal{O}_n'^{EM}(\Lambda) | H \rangle, \quad (2)$$

for the electromagnetic decays, where $\Lambda$ is a factorization scale and $f_n$'s are the coefficients from four-fermi vertices.

For the case of bottomonium decays, we have calculated the lattice regulated matrix elements in this factorization theorem[2], using the leading order non-relativistic QCD lagrangian

$$\mathcal{L} = \psi^\dagger (D_t - \frac{\vec{D}^2}{2M}) \psi + \chi^\dagger (D_t + \frac{\vec{D}^2}{2M}) \chi. \quad (3)$$

The matrix elements in which we are interested are :

$$G_1 = \langle {}^1S | \psi^\dagger \chi \chi^\dagger \psi | {}^1S \rangle / M^2 \quad (4)$$

*Work supported by the U.S.Department of Energy, Division of High Energy Physics, Contract W-31-109-ENG-38. Work done in collaboration with D.K.Sinclair and G.T.Bodwin
†current address : Center for Theoretical Physics, Seoul National University, Seoul,Korea

$$F_1 = \langle {}^1S | \psi^\dagger \chi | 0 \rangle \langle 0 | \chi^\dagger (-\frac{i}{2} \overleftrightarrow{D})^2 \psi | {}^1S \rangle / M^4 \quad (5)$$

$$H_1 = \langle {}^1P | \psi^\dagger (i/2) \overleftrightarrow{D} \chi \chi^\dagger (i/2) \overleftrightarrow{D} \psi | {}^1P \rangle / M^4 \quad (6)$$

$$H_8 = \langle {}^1P | \psi^\dagger T^a \chi \chi^\dagger T^a \psi | {}^1P \rangle / M^2. \quad (7)$$

To make use of lattice regulated matrix elements (which we have calculated already in [2]) for phenomenology, we need to perform operator "renormalization" which translates lattice regulated matrix elements into continuum $\overline{MS}$ matrix elements,

$$\mathcal{O}_i^L = \sum_j Z_{ij} \mathcal{O}_j^{\overline{MS}}, \quad (8)$$

where $\mathcal{O}$'s are ordered following the velocity scaling law [3]. Once we calculate the matrix $Z_{ij}$, we obtain $\mathcal{O}_i^{\overline{MS}}$ by inverting the Z matrix :

$$\mathcal{O}_i^{\overline{MS}} = \sum_j Z_{ij}^{-1} \mathcal{O}_j^L. \quad (9)$$

From $G_1, F_1, H_1$, and $H_8$, we note that necessary $\mathcal{O}$'s are :

$$I \equiv \psi^\dagger \chi \chi^\dagger \psi, \quad (10)$$

$$D^2 \equiv \psi^\dagger \chi | 0 \rangle \langle 0 | \chi^\dagger (-\frac{i}{2} \overleftrightarrow{D})^2 \psi \quad (11)$$

for the S-wave case and

$$h_1 \equiv \psi^\dagger (i/2) \overleftrightarrow{D} \chi \chi^\dagger (i/2) \overleftrightarrow{D} \psi, \quad (12)$$

$$h_8 \equiv \psi^\dagger T^a \chi \chi^\dagger T^a \psi, \quad (13)$$

for P-wave. Here, we dicuss our preliminary results on the renormalization of these operators (full details will be reported elsewhere [4]).

Firstly, we derive Feynman rules for our lattice lagrangian upto $\mathcal{O}(g^2)$ from the lattice quark propagator,

$$G(\vec{x}, t+1) = (1 - H_0/2n)^n U^\dagger(\vec{x}, t)(1 - H_0/2n)^n G(\vec{x}, t)$$
$$+ \delta_{\vec{x},\vec{0}}\delta_{t+1,0}, \qquad (14)$$

with $G(\vec{x}, t) = 0$ for $t < 0$ and $H_0 = -\Delta^2/2M_0 - E_{sub}$ ($\Delta$ is lattice covariant derivative, $E_{sub} = 3(1-u_0)/M_0$, $u_0 = \langle 0|\frac{1}{3}\text{Tr}U_{plaq}|0\rangle^{\frac{1}{4}}, n = 2$). Then, using these Feynman rules, we calculate one loop renormalization of the operators, (10) – (13).

Necessary Feynman diagrams for the part of $D^2$ which contributes to the mixing with the continuum operator $I$ are given in Fig 1. A dotted line means the temporal component of the gluon field and a wavy line means the spatial component of gluon field (in the following figures, we use the same convention). a) has the lattice $\nabla^2$ vertex, b) has the $g^2 \vec{A}^2$ vertex, and c) is the tadpole contribution. After finding analytic expressions for these diagrams, we take $\vec{p}$ (external momentum) $\to 0$(except the tadpole contribution) to pick out pieces for the continuum $I$.

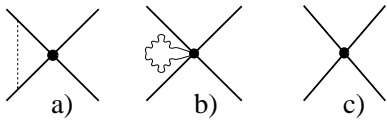

Figure 1. Feynman diagram for the power divergent part of $D^2$ renormalization.

Similarly, Fig 2 is the Feynman diagrams which need to be evaluated for the finite renormalization of $D^2$ operator. a) and b) have the lattice $\nabla^2$ vertex, c) has $\chi^\dagger \psi$ vertex, d) is the tadpole contribution, e) has the $g\vec{A} \cdot \vec{\nabla}$ vertex, and f) has the $g^2 \vec{A}^2$ vertex. We take $\frac{\partial^2}{\partial p_i^2}$ (analytic expression for these diagrams) $|_{\vec{p}=0}$ in order to get the finite renormalization of $D^2$ except for the case of the tadpole contribution.

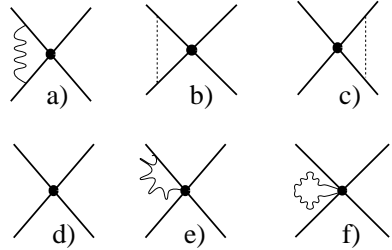

Figure 2. Feynman diagram for the multiplicative renormalization of $D^2$ operator.

The finite renormalization of the operator, $I$, is given by the Feynman diagrams of Fig 3. Both a) and b) have $\chi^\dagger \psi$ vertex. Again, to get the finite renormalization of $I$, we set $\vec{p} = 0$ in the analytic expression for the Feynman diagrams. Expansion of these Feynman digrams in terms of the external momentum tells us that this operator mixes with continuum $\nabla^2$ at order $\mathcal{O}(\alpha_s v^2)$ (actually, both lattice $D^2$ and $I$ mix with all continuum terms in arbitrary powers of $v^2$). However, since $\alpha_s \sim v^2$, this effect is in higher order of $\alpha_s$ and we neglect such mixing effect to be consistent with the order we are calculating.

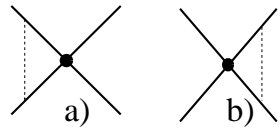

Figure 3. Feynman diagram for the multiplicative renormalization of $I$ operator.

Fig 4 shows the Feynman diagrams for the finite renormalization of the operator, $h_8$. The vertex factor is $T^a \otimes T^a$. Expansion in terms of the external momentum shows that this term does not mix with $h_1$. With $\vec{p} = 0$, we obtain the finite renormalization contribution from these diagrams.

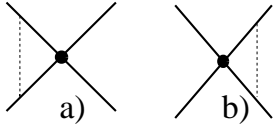

Figure 4. Feynman diagrams the multiplicative renormalization of $h_8$

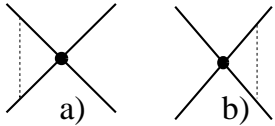

Figure 5. Feynman diagrams for the multiplicative renormalization of $h_1$

With $\vec{p} = \vec{p}'' = 0$, the Feynman diagrams in Fig 5 gives the contribution to the finite renormalization of the operator, $h_1$. The vertex factor is lattice version of $\vec{\nabla} \otimes \vec{\nabla}$.

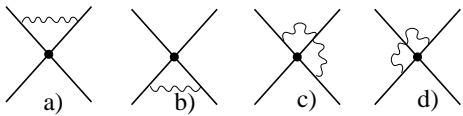

Figure 6. Feynman diagrams for the mixing contribution from $h_1$ to $h_8$

Finally, Fig 6 shows the Feynman diagrams for the mixing of the operator, $h_1$, with $h_8$. The vertex factor is the same as Fig 5. With zero external momentum, these diagrams have a logarithmic infrared divergence. Therefore, these digrams need extra care. We subtract logarithmic divergent piece from the analytic expression of these diagrams if $\sum_i p_i^2 \leq \pi^2$ when the integral is evaluated numerically. The last entry in Table 1 is the difference between the logarithmically divergent piece which we use in the subtraction and that for the $\overline{MS}$ expression.

Table 1
numerical values of Feynman integrals

| diagram | Vegas output |
|---|---|
| fig 1a | 0.4781(8) |
| fig 1b | 0.7008(2) |
| fig 1c | -0.8448(100) |
| fig 2a | 0.12476(3) |
| fig 2b | -0.18774(3) |
| fig 2c | 0.005360(4) |
| fig 2d | 0.1411(100) |
| fig 2e | -0.27241(5) |
| fig 2f | -0.116804(7) |
| fig 3 | 0.010717(2) |
| fig 4a+4b | -0.001346(3) |
| fig 5a+5b | -0.11773(3) |
| fig 6a+6b+6c+6d | -0.0060293(9) |
| (fig 6a+6b+6c+6d) log | 0.0013 |

After we get the analytic expression for the Feynman diagrams we listed in the above, we evaluate the integrals numerically by use of VEGAS. Table 1 is the summary of the numerical values of each Feynman integral. We used $\alpha_s = 0.135$ and $M_b = 1.71$ in lattice units. Note that $M_b$ in the integral differs from that in the simulation (1.5) because the tadpole improvement scheme we employed tells us that $M_b = M_0/u_0$.

For the S-wave, from the relations,

$$D_{Lat}^2 = (1+G)D^2 + FI \qquad (15)$$
$$I_{Lat} = (1+E)I, \qquad (16)$$

where $G = $ (sum of fig 2 contributions), $F = $ (sum of fig 2 contributions), and $E = $ (sum of fig 3 contribution), we get

$$D^2 = \frac{1}{(1+G)} D_{Lat}^2 - \frac{F}{((1+E)(1+G))} I$$

$$I = \frac{1}{(1+E)} I_{Lat}. \tag{18}$$

For the P-wave, from the relations,

$$\Delta_{Lat} \otimes \Delta_{Lat} = (1+A)\Delta \otimes \Delta \tag{19}$$

$$T^a_{Lat} \otimes T^a_{Lat} = (1+H)T^a \otimes T^a + J\Delta \otimes \Delta, \tag{20}$$

where $A$ = (sum of fig 5 contributions), $J$ = (sum of fig 6 contributions), and $H$ = (sum of fig 5 contributions), we obtain

$$\Delta \otimes \Delta = \frac{1}{(1+A)} \Delta_{Lat} \otimes \Delta_{Lat} \tag{21}$$

$$T^a \otimes T^a = \frac{1}{(1+H)} T^a_{Lat} \otimes T^a_{Lat} - \frac{J}{((1+H)(1+A))} \Delta_{Lat} \otimes \Delta_{Lat}. \tag{22}$$

From [2], we recall $G_1^{Lat} = 0.1488(5)\frac{1}{M_b^2}$, $F_1^{Lat} = 1.3134(9)\frac{G_1}{M_b^2}$, $H_1^{Lat} = 0.0145(6)\frac{1}{M_b^4}$, $H_8^{Lat} = 0.0152(3) M_b^2 H_1$. Thus, after putting all the coefficients together, we get $M_b^2 G_1^{\overline{MS}} = 0.147$, $M_b^4 F_1^{\overline{MS}} = 0.133$, $M_b^4 H_1^{\overline{MS}} = 0.01641$, $M_b^2 H_8^{\overline{MS}} = 0.000337$ in lattice units (without error bars). With these $\overline{MS}$ matrix elements, we estimate the decay rates for various processes using formulae in [1]. With $\alpha_s(M_b) = 0.20, M_b = 4.7\text{GeV}, \alpha_{em} = 1/128$, we get : For the S-wave case,

$$\begin{aligned}\Gamma(\Upsilon \to LH) &= [\frac{(\pi^2-9)(N_c^2-4)}{54 N_c} C_F \alpha_s^3(M_b) \\ &\quad \{1 + (-9.46(2) C_F \\ &\quad + 4.13(17) C_A - 1.161(2) n_f) \frac{\alpha_s}{\pi}\} \\ &\quad + \pi Q^2 (\sum_i Q_i^2) \alpha_{em}^2 (1 - \frac{13\alpha_s}{4\pi} C_F)] 2 G_1 \\ &= 36\text{KeV}\end{aligned}$$

$$\begin{aligned}\Gamma(\Upsilon \to l^+ l^-) &= \frac{2\pi}{3} Q^2 \alpha_{em}^2 (1 - \frac{16\alpha_s(M_b)}{3\pi}) G_1 \\ &= 0.98\text{KeV}\end{aligned}$$

For P-wave case,

$$\begin{aligned}\Gamma(^1P_1 \to LH) &= \frac{\pi n_f}{3} \alpha_s^2(M_b) H_8 = 26\text{KeV}\end{aligned}$$

$$\begin{aligned}\Gamma(^3P_0 \to LH) &= \pi \alpha_s^2(M_b)(C_F H_1 + \frac{n_f}{3} H_8) \\ &= 380\text{KeV}\end{aligned}$$

$$\begin{aligned}\Gamma(^3P_1 \to LH) &= \frac{n_f \pi}{3} \alpha_s^2(M_b) H_8 = 26\text{KeV}\end{aligned}$$

$$\begin{aligned}\Gamma(^3P_2 \to LH) &= \pi \alpha_s^2(M_b)(\frac{4}{5N_c} C_F H_1 + \frac{n_f}{3} H_8) \\ &= 120\text{KeV}\end{aligned}$$

$$\Gamma(^3P_0 \to \gamma\gamma) = 6\pi Q^4 \alpha_{em}^2 F_1 = 0.26\text{KeV}$$

$$\Gamma(^3P_2 \to \gamma\gamma) = \frac{8\pi}{5} Q^4 \alpha_{em}^2 F_1 = 0.070\text{KeV}$$

Among these decay rates, experimental values exist only for electromagnetic and hadronic decay of the S-wave $\Upsilon$. In this case, we can compare the lattice values (36KeV, 0.98KeV)) to the experimental values (48KeV, 1.3KeV) and we find the agreement is reasonable.

The remaining work is the scale setting for each integral listed in the above and the work is currently in progress [4].

**Acknowlegement**

S.Kim would like to thank prof.'s H.S. Song and C. Lee of Center for Theoretical Physics at Seoul National University for their hospitality.

**REFERENCES**

1. G.T.Bodwin, E.Braaten, and G.P.Lepage, Phys. Rev. **D51** (1995) 1125.
2. G.T.Bodwin, S.Kim, and D.K.Sinclair, Nucl. Phys. **B**(Proc. Suppl.) 34 (1994) 347.
3. G.P.Lepage, L.Magnea, C.Nakhleh, U.Magnea, and K.Hornbostel, Phys. Rev. **D46** (1992) 4052.
4. G.T.Bodwin and S.Kim, work in progress.